\documentclass[%
superscriptaddress,
amsmath,amssymb,
aps,
prfluids,
preprint,
]{revtex4-2}
\usepackage{siunitx}
\usepackage{ctable}
\usepackage{threeparttable}
\usepackage{diagbox}
\usepackage{float}
\usepackage{overpic}
\usepackage{epstopdf}
\usepackage{graphicx}
\usepackage{dcolumn}
\usepackage{bm}
\usepackage{epsfig,dsfont,amssymb,amsmath,amsthm,amsfonts,amsbsy,mathrsfs}
\usepackage{siunitx}
\usepackage{url}

\usepackage{hyperref} 
\hypersetup{hypertex=true,
	colorlinks=true,
	linkcolor=blue,
	anchorcolor=blue,
	citecolor=blue}

\begin{document}
	
\title{Emergence of a hexagonal pattern in shear-thickening suspensions under orbital oscillations}
\author{Li-Xin Shi}
\author{Meng-Fei Hu}
\author{Song-Chuan Zhao}
\email[]{songchuan.zhao@outlook.com}
\affiliation{State Key Laboratory for Strength and Vibration of Mechanical Structures,\\ School of Aerospace Engineering, Xi’an Jiaotong University, Xi’an 710049, China}
\begin{abstract}
Dense particle suspension under shear may lose its uniform state to large local density and stress fluctuations, which challenge the mean-field description of the system. Here, we explore the novel dynamics of a non-Brownian suspension under orbital oscillations, where localized density waves along the flow direction appear beyond an excitation frequency threshold and self-organize into a hexagonal pattern across the system. The spontaneous occurrence of the inhomogeneity pattern arises from a coupling between particle advection and the shear-thickening nature of the suspension. Through linear stability analysis, we show that they overcome the stabilizing effects of particle pressure at sufficient particle volume fraction and oscillation frequency. In addition, the long-standing density waves degenerate into random fluctuations when replacing the free surface with rigid confinement. It indicates that the shear-thickened state is intrinsically heterogeneous, and the boundary conditions are crucial for developing local disturbance.
\end{abstract}
\maketitle
\section{Introduction}
\label{sec:headings}

Dense suspensions composed of mixtures of particles and fluid are ubiquitous in natural phenomena and industrial processes~\citep{Larson1998,wagner2009shear}. For a sufficiently large volume fraction of particles, $\Phi$, the suspensions can exhibit a wide range of nonlinear phenomena, including yielding, shear thinning or thickening, and shear jamming ~\citep{guazzelli2011physical,morris2020shear,nabizadeh2022structure}, which has stimulated decades of research in the rheology and physics community ~\citep{wagner2009shear,guazzelli2018rheology}. In particular, discontinuous shear thickening (DST), where the suspension viscosity, $\eta$, increases over several orders of magnitude, has recently been understood as a transition from lubricated to frictional particle interactions when the applied stress overcomes the interparticle repulsion~\citep{Fernandez2013,wyart2014discontinuous,seto2013discontinuous, Brown_2014}. Readers may refer to \citet{morris2020shear} for a recent review. The constitutive curve of such a suspension, relating the shear stress $\tau$ and the shear rate $\dot{\gamma}$, displays a $\mathsf{S}$-shape in which a negatively sloped region connects the lubricated branch of low-viscosity at low stresses and frictional branch of high-viscosity at high stresses~\citep{Brown_2014,mari2015nonmonotonic, Pan2015}. Because of the characteristic $\mathrm{d} \dot{\gamma} / \mathrm{d} \tau\leq 0$, the steady uniform state of a suspension under shear may become unstable and reduced to large spatiotemporal fluctuations of stress and densities~\citep{lootens2003giant, rathee2017localized, rathee2020localized}. 
Structures and patterns, such as \textit{gradient banding} \citep{hu1998shear,olmsted2008perspectives} and \textit{vorticity banding}~\citep{saint2018uncovering,chacko2018dynamic}, may appear as a consequence.
Those inhomogeneities, either transient or periodic~\citep{hermes2016unsteady, ovarlez2020density}, challenge the mean-field description of a shear-thickening suspension.
Nevertheless, little is known about the disturbance growth in shear thickening suspensions in flow configurations beyond simple shear~\citep{darbois2020surface, Rathee2021}. Moreover, although boundary confinements are proposed as essential for the shear-thickening of non-Brownian suspensions~\citep{Brown_2014}, their influence on the instability development in such systems, if any, have yet to be explored.

This paper investigates the instability arising in the \textit{flow} direction in dense Non-Brownian suspensions under orbital shaking. A novel unsteady dynamics where density waves self-organize into a hexagonal pattern is observed, accompanied by large spatial stress fluctuations. Our result shows that it is closely related to the shear thickening of the suspension. Moreover, the essential role of boundary conditions in the development of heterogeneity is discussed.

\section {Experimental protocol}
\begin{figure*}
	\centering		
	\includegraphics[width=\textwidth]{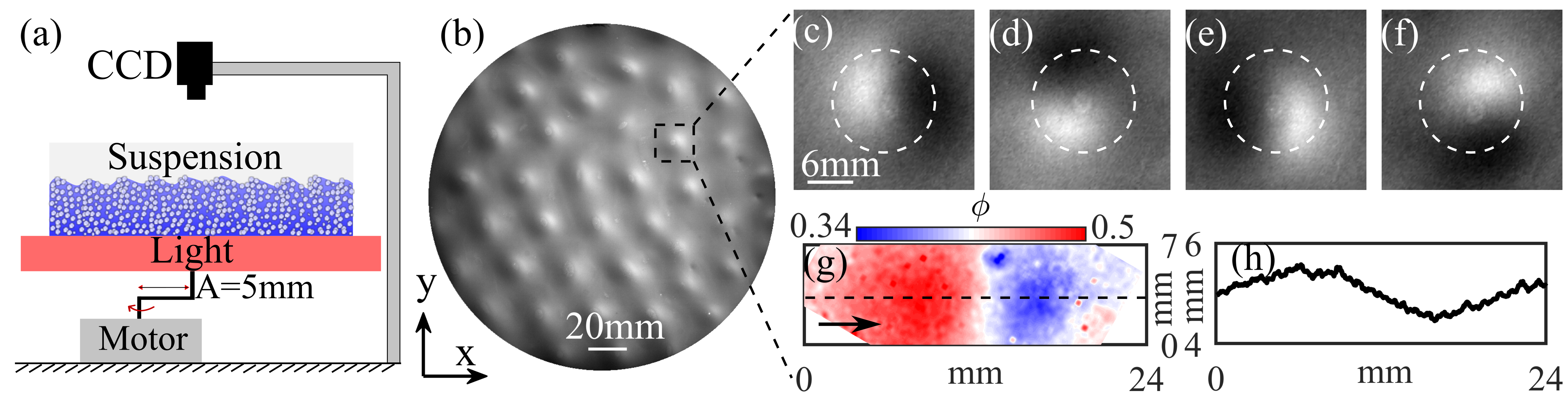}
	\caption{(a) A schematic of the experimental setup; (b) A top-view snapshot of the suspension at $f = 7.67 \ \si{Hz}$ (Movie 1). The dark  (bright) areas correspond to the high (low) $\phi$ regions. (c-f) A zoomed area at different phases in one oscillation period. The dashed circle denotes the trajectory of the center of the dense area. An example of local particle fraction, $\phi$, measurement is shown in (g). The arrow indicates the instantaneous flow direction. Note that $\phi$ is computed accounting for both the light transmission ratio and the surface deformation. Panel (h) illustrates the surface height along the dashed line in (g). }
	\label{fig:uniform_densitywave}
\end{figure*}

The suspension consists of deionized water mixed with cornstarch particles of an average size $d=15 \ \si{\mu m}$, confined within an open cylindrical container. The mass density of dry starch particles, $\rho_p$, is 1.61\si{kg/m^3}. When preparing density-matched suspensions, cesium chloride is added to the solution.
The container is subjected to a horizontal orbital vibration, i.e., a circular translational motion of the entire platform. The oscillation frequency, $f$, can be tuned from $0.5$ to $8.33~\si{Hz}$ with a fixed amplitude $A = 5~\si{mm}$ indicating the radius of the orbital motion [Fig.~\ref{fig:uniform_densitywave}a]. 
An acrylic plate can be optionally laid on the surface of the suspension, either moving with the container or fixed in the lab reference. The following experiments are performed in the free surface configuration, unless stated otherwise. 
A monochrome LED panel from below illuminates the suspension. A high-speed camera (Microtron EoSens 1.1cxp2) is fixed in the laboratory frame of reference, recording the light transmission. The spatial resolution is 0.23\si{mm} per pixel. The local thickness of suspension, $h$, can be measured on demand via an in-house-built laser profilometry~\citep{Zhao2015}, and the typical resolution of $h$ is 61\si{\micro\meter/pixel} in the current configuration. 
The transmission of the bottom light is through scattering events. We established a one-dimensional multi-scattering model, and the ratio between the incoming flux and the transmitted flux is given as $F(\phi,h) = 1 + a \phi^bh$. The parameters $a$ and $b$ are calibrated with the light intensity attenuation data for uniform cornstarch suspensions. Readers may find the calibration data and method justification in Appendix~\ref{appA}. 
By simultaneously measuring $h$ and $F(\phi,h)$, the local volume
fraction, $\phi$, can be calculated. An example is given in Fig.~\ref{fig:uniform_densitywave}g-h, where both $h$ and $\phi$ are measured. 
The container has a diameter of $32 \ \si{cm}$ and a height of $5 \ \si{cm}$. Though the side wall is critical for the surface wave of a Newtonian fluid of low viscosity, such as water, under orbital shaking~\citep{Reclari2014}, its effect is ignored here, as changing the diameter of the container from 32 $\si{cm}$ to 16 $\si{cm}$ does not alter the phenomena studied in this work. 
The system is then defined by three major experimental parameters: the oscillation frequency $f$, the global particle packing fraction $\Phi$, and the average thickness of the suspension $h_0$. We first report our main observations in a reference system of $\Phi = 0.42$ and $h_0 = 5 \si{mm}$. The influence of $\Phi$ and the top confinement will be discussed afterward. 

\section {Results}
\subsection{Experimental Results}
The suspension is uniform at low oscillation frequencies. However, when $f$ is increased beyond $3.67 \ \si{Hz}$, high-density regions appear suddenly, followed by low-density tails. These density patches self-organize into a hexagonal pattern with a further increase of $f$, as shown in Fig.\ref{fig:uniform_densitywave}(b). Individual dense regions move in a circular path at the same frequency $f$ precisely, with little changes in location and size over time (Fig.~\ref{fig:uniform_densitywave}c-f; refer to supplementary materials for multimedia view). The motion of the dense regions is defined by the position of the highest density locally. The typical diameter of the circular path, denoted as $d_w$, measures 13.3 \si{mm}, which is larger than the orbit of the container. In such a state, the surface becomes uneven too, introducing a peak-to-valley height difference $\Delta h\approx 1\si{mm}$. Dense/loose regions correspond to peaks/valleys [Fig.~\ref{fig:uniform_densitywave}g]. As the suspension is bounded by the air-liquid interface, the observation indicates non-uniform normal stress distribution accompanying the density inhomogeneity, a hallmark of non-Newtonian behavior. 

\begin{figure}
	\centerline{\includegraphics[scale=0.9]{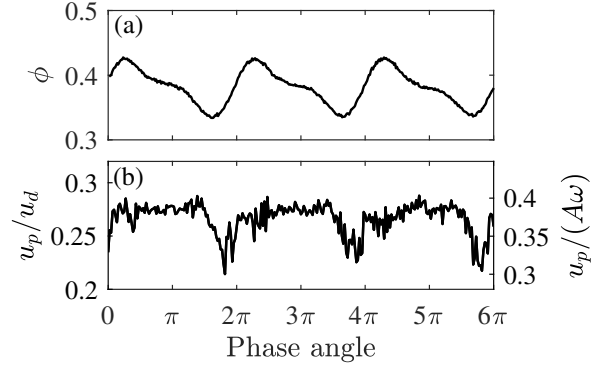}}
	\caption{The local volume fraction $\phi$ (a) and particle velocity $u_p$ (b) varies with the phase angle of the oscillation. Both $\phi$ and $u_p$ rise when a density wave passes through the measuring area. $u_w=\pi d_w f$ is the propagation velocity of the density wave concerned. Date here is collected at $\Phi=0.41$, $f=5.83\si{Hz}$ and $h=3\si{mm}$. The shallower suspension used here, thus more subtle surface deformation, improves the accuracy of the measurement of $\phi$ and $u_p$.}
	\label{fig:particle_vel} 
\end{figure}

As reported by \citet{galvez2017dramatic}, the interaction force profile shows a hysteresis when corn starch particles are pressed into contact. However, the inhomogeneity transition here is reversible, which implies that the dense regions in our experiments are not permanent aggregates of contacting particles. To reveal the microscopic nature of the observed inhomogeneity, we measure the velocity of tracer particles near the suspension surface, $u_p$, and local packing fraction, $\phi$, simultaneously at a point on the path of a high-density region. Both $u_p$ and $\phi$ vary in phase with the same period as the oscillation, as shown in Fig.\ref{fig:particle_vel}, \textit{i.e.,} particles in denser regions tend to move faster. However, the particle velocity remains lower than that of the density pattern, indicating that the motion of the observed density pattern is the propagation of density waves. Furthermore, the gradient of $\phi$ in the wavefront is significantly steeper than that at the rear [see Fig.\ref{fig:uniform_densitywave}(g)], a signature of shock waves. 

\subsection {Instability onset}
\label{sec:onset}
The non-uniform state can not be modeled using a single-phase description of the suspension. In the exploration of flow instabilities in particulate suspensions, two-fluid models are normally used~\citep{chacko2018dynamic, Batchelor1988}. 
The continuity equation of particles reads
\begin{equation}
	\frac{\partial \phi}{\partial t}+\nabla \cdot(\vec{u}_p\phi)=0.
	\label{eq.p_continue}
\end{equation}
When the inertia of particles is neglected, Eq.~\ref{eq.p_continue} can be written in the form of an advection-diffusion equation of $\phi$~\citep{Anderson1995}. Shock waves are thus expected in certain circumstances. However, it has been demonstrated that the inertia of particles (the advective term) is critical for instability development~\citep{Batchelor1988,Johri2002}. Therefore, we consider the momentum equation of particles in addition.

The general analytical formulation of the stress tensor of the particle phase remains a difficult task to tackle after decades of efforts~\citep{Jackson2000,guazzelli2018rheology}. Here, we exclusively consider the dominant terms  for relatively dense suspensions:
\begin{equation}
	\phi \rho_{p} \left( \frac{\partial{\vec{u}_p}}{\partial t}+(\vec{u}_p \cdot\nabla)\vec{u}_p \right) =C_d(\vec{U}-\vec{u}_p)+\eta_p\nabla^{2}\vec{u}_p-\nabla \Pi.
	\label{eq.p_momentum}
\end{equation}
On the right-hand side of Eq.~\ref{eq.p_momentum}, the first term represents the hydrodynamic drag exerted on particles, proportional to the relative velocity between the particle $\vec{u}_p$ and the average flow of the mixture $\vec{U}$, and the Richardson-Zaki approximation for $C_d$~\citep{Richardson1954,Buscall1982} is used. Note that the mixture flow $\vec{U}$ is defined as the weighted average of the velocity of the two phases. The second term is the viscous force between particles, where $\eta_p$ is the dynamic viscous coefficient. The third term describes the gradient of particle pressure, $\Pi$. Both $\eta_p$ and $\Pi$ will be evaluated by their suspension counterparts, as the stresses of particle phase dominate the suspension dynamics at high $\phi$~\citep{Gallier2014}. The stability analysis of Eqs.~\ref{eq.p_continue}-\ref{eq.p_momentum} will be performed with respect to the uniform state. Therefore, the mixture velocity $\vec{U}$ of the suspension in the uniform state is computed first.

Consider a suspension with a volume fraction $\Phi$ that behaves as a uniform fluid with a kinematic viscosity $\nu$ and is subjected to bottom oscillation. Since the orbital motion is the superposition of two perpendicular harmonic oscillations of angular frequency $\omega=2\pi f$ with a phase difference of $\pi/2$, the flow velocity $\vec{U}=(U_x,\,U_y)$ is written as a complex function $U(z,t) = U_x + i U_y$ which satisfies
	\begin{subequations}
		\label{eq.s_stokes}
		\begin{align}
			\frac{\partial U}{\partial t} &= \nu \frac{\partial^{2}{U}}{\partial z^{2}}\\
			\frac{\partial{U}}{\partial z}\Big\vert_{z=h} = 0 & \quad\mathrm{and }\quad
			U(z=0,t) = A\omega e^{i\omega t}. 		
		\end{align}
	\end{subequations}  
Equation~\ref{eq.s_stokes}~b indicates the no-slip condition on the bottom and the zero-shear condition on the free surface. 
For shear-thickening fluid, the effective kinematic viscosity of the suspension, $\nu = \nu_{0}(\phi_{J}-\phi)^{-2}$, is rate-dependent~\citep{guazzelli2018rheology,boyer2011unifying,singh2018constitutive}.
A recently developed phenomenological constitutive model~\citep{wyart2014discontinuous}, based on the mean-field description of shear-thickening process, suggests that $\phi_J$ experiences a crossover from its frictionless value $\phi_{0}$ to $\phi_{m}$ for frictional contacts, when the typical stress $\tau$ in the suspension exceeds a characteristic value $\tau^*$, \textit{i.e.,}  $\phi_{J}=\phi_{0}-e^{{-\tau^{*}}/{\tau}}(\phi_0-\phi_m)$. For the sample used here, we found $\nu_0=0.95 \times 10^{-6}~\si{m^2/s}$, $\tau^{*}=3.7~\si{Pa}$, $\phi_m=0.45$ and $\phi_0=0.58$ according to rheology measurements (see Appendix~\ref{sec:rheo}).

Analogous to Stokes problem in two dimensions, the solution of Eq.~\ref{eq.s_stokes} can be described by a dimensionless number $l=\sqrt{{2\nu}/{\omega}}/{h}$~\citep{yih1968instability}, the square of which is merely the inverse Reynolds number representing the significance of viscosity relative to inertia. For $l\sim 1$, $U$ follows the orbital motion of the bottom plate with little phase lag and magnitude decay along with $z$. 
Due to the rate-dependence of $\nu$, the shear stress $\tau=\rho\nu\vert\partial U/\partial z\vert$ and $l$ are interrelatted for a given $\phi=\Phi$ (details are provided in Appendix~\ref{appB}), where $\rho$ is the suspension density.
To simplify the calculation, $\tilde{U} = \vert U(z=h,t)\vert$ and $\tilde{\tau} = \tau(z=0)$ are used to characterize the mainstream flow of the suspension in the uniform state. 
We assume that the initial development of the instability occurs in the mainstream direction, which results in the observed motion of density patterns. This assumption is significant, as  Eqs~\ref{eq.p_continue}-\ref{eq.p_momentum} are hence reduced to one dimension (the main flow direction), and particle migration along the gradient and the vorticity directions are neglected. A linear stability analysis can be readily performed for the reduced equations. We leave the calculation details in Appendix~\ref{appB} and present the main result here.
The uniform flow ($\phi=\Phi$ and ${u_p}=\tilde{U}$) becomes unstable against density perturbations of a wavenumber $k$, provided~\citep{zhao2021spontaneous} 
\begin{equation}
	\mathcal{C}+k^2 \frac{\eta_p}{C_d}<\frac{\phi \tilde{U}^{\prime}}{\sqrt{\Pi^{\prime}/\rho_p}},
	\label{eq.instability}
\end{equation}
where $\tilde{U}^{\prime}$ and $\Pi^{\prime}$ are the derivatives of $\tilde{U}$ and $\Pi$ with respect to $\phi$ at $\phi=\Phi$.  In theory, $\mathcal{C}= 1$ is a constant whose value is to be adjusted by comparing with experiments, accounting for the neglected features in the model. 

\begin{figure}
	\centerline{\includegraphics{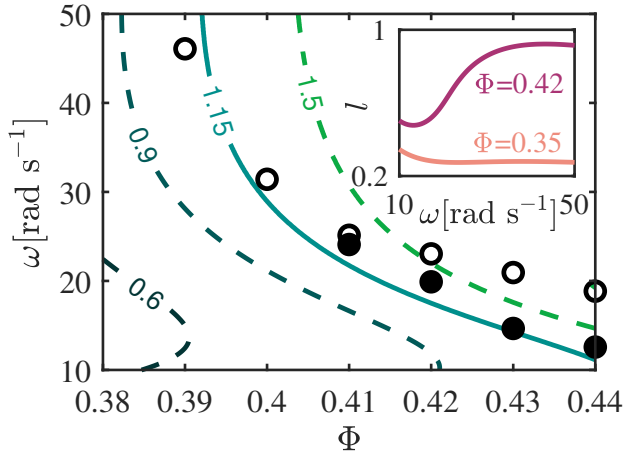}}
	\caption{State diagram. The onset frequency $\omega_c$ of density waves is denoted by open (solid) circles for the (density-matched) aqueous cornstarch suspension. The contour lines of $\Phi \tilde{U}^{\prime}/{\sqrt{\Pi^{\prime}/\rho_p}}$ (the right-hand side of Eq.~\ref{eq.instability}) are plotted for comparison. Inset: The dimensionless number $l$ varies with $\omega$ for $\Phi =0.35$ (no DST) and $\Phi =0.42$ (DST occurs) respectively.}
	\label{fig:State_diagram}
\end{figure}

It is clear in Eq.~\ref{eq.instability} that the viscosity of the particle phase, $\eta_p$, stabilizes the short-wave disturbances. Therefore, the instability first occurs at the long wave limit ($k\sim 0$). Denoting the container diameter as $L=\SI{32}{cm}$, the second term on the left-hand side, $\sim \eta_0/C_d L^2 \sim d^2/L^2 \sim \mathcal{O}(10^{-5})$, is negligible.
The onset of instability is thus insensitive to the value of $\eta_p$, which, however, shapes the developed density waves (see discussion on Fig.~\ref{fig:wavenumber} below). On the right-hand side of Eq.~\ref{eq.instability} $\tilde{U}^\prime$ introduces a shock-wave like kinematic instability promoting the growth of high $\phi$ regions, \textit{i.e.,} a higher $\phi$ leads to a faster flow locally, as confirmed in Fig.~\ref{fig:particle_vel}. The particle pressure term $\Pi^{\prime}$, on the other hand, provokes the particle migration out of the local high $\phi$ region. The competition between these two terms sets the onset of instability. Here, $\Pi = \tilde{\tau}(l,\phi)$ is used~\citep{brown2012role}. With all terms defined, we calculate the onset frequency of the disturbance growth, $\omega_c$, and find that $\mathcal{C}=1.15$ aligns with the experimental observation [Fig.~\ref{fig:State_diagram}]. In particular, the theoretical $\omega_c(\Phi)$ agrees with its experimental counterparts quantitatively for the density-matched suspensions, where buoyancy is absent. Otherwise, the onset frequency $\omega_c$ is obscured by the suspending threshold of the relatively dense suspension without density-match in practice. The resultant higher value of $\mathcal{C}$ than the theory may be explained by the neglect of the particle migration in the vorticity direction and the implemented equality of $\Pi=\tilde{\tau}$. Both underestimate the denominator of the right-hand side of Eq.~\ref{eq.instability}.

The model predicts a minimum packing fraction around $\Phi=0.392$ [Fig.~\ref{fig:State_diagram}], below which Eq.~\ref{eq.instability} can not be satisfied, and the uniform state is always stable. This result is closely related to the shear-thickening nature of the suspension and reveals the underlying mechanism of the instability development. The solution for $\tilde{U}$ in Eq.~\ref{eq.s_stokes} increases rapidly at intermediate $\phi$, thus corresponding to a regime of large $\tilde{U}^\prime$. An increase of $l$ would shift this regime toward lower densities. The pressure term $\Pi$ follows a similar trend. It can be shown that the variation of $\Phi \tilde{U}^{\prime}/{\sqrt{\Pi^{\prime}/\rho_p}}$ is dominated by $\sqrt{ l/(\phi_J-\Phi)}$ for $\omega\gtrsim 10~\si{rad~s^{-1}}$ (see Appendix~\ref{appB}). The effective viscosity of the suspension $\nu$ in general grows with $\omega$, {i.e.,} $\nu\sim\omega^\alpha$ and $\alpha>0$. When approaching DST, $\alpha$ becomes considerably larger than 1, and $l\sim\omega^{(\alpha-1)/2}$ displays a dramatic increase.  
Figure~\ref{fig:State_diagram} inset illustrates this distinct behavior. A minimum $\Phi$ is associated with the occurrence of DST ~\citep{hermes2016unsteady,fall2015macroscopic}, thus the same holds for the observed density waves here. The relation with shear-thickening is additionally confirmed via experiments of an aqueous solution of polydisperse silica beads (of an average diameter of 20 $\mu \si{m}$) in the same setup. It was known that dissolving electrolytes within the solvent reduces the magnitude of repulsive forces between silica grains due to surface charge, and DST disappears accordingly~\citep{clavaud2017revealing}. We confirm that the density waves in the suspension of silica beads under orbital oscillations are significantly weakened in the same manner (Movie 3 and 4).

\subsection{Development of density waves}
\begin{figure}
	\centerline{\includegraphics[width=8.6cm]{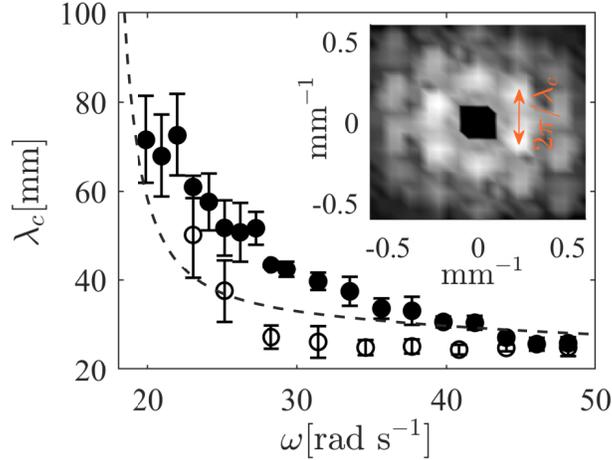}}
	\caption{The lengthscale of the observed density pattern $\lambda_c$ varies with $\omega$ in the aqueous-starch suspension (open circles) and the density-matched suspension (solid circles). Experimental parameters here are  $\Phi=0.42$, $h=5\si{mm}$. The dashed line is the wave number calculated by the linearized model with a proliferated viscosity, $\lambda_m$. See the main text for discussions. Inset: two-dimensional Fourier spectrum of Fig.\ref{fig:uniform_densitywave}(b). }
	\label{fig:wavenumber}
\end{figure}

Once the disturbance arises and grows in the flow direction, secondary instabilities may further develop~\citep{Anderson1995,Duru2002}, and two-dimensional structures form~\citep{Glasser1998}. In our experiments, the density waves self-organize into a hexagonal pattern as $f$ increases (cf. Fig.~\ref{fig:uniform_densitywave} and~\ref{fig:wavenumber} Inset). A comprehensive theoretical analysis of the formation of the observed pattern is beyond the scope of this work. Instead, we argue that it can be understood from the symmetry perspective. Consider that the uniform state becomes unstable at one moment and breaks into alternating density bands perpendicular to the flow direction and separated by a characteristic wavelength $\lambda_c$. In the following moments, however, the flow direction changes constantly in the horizontal plane. Those bands can not preserve the alignment with the flow without breaking the translational symmetry in the lateral (vorticity) direction. Therefore, the band structure is unstable and must be reduced to localized density patches. Particle migration dominates when those high-density regions are misaligned. The most stable structure thus maximizes the duration of the alignment with the flow. In other words, the favorite structure retains the highest rotational symmetry and a discrete translational symmetry of $\lambda_c$ in two-dimension, \textit{i.e.,} the hexagonal pattern, as observed. A subtle inference along this line of argument is that the maximum local density fluctuates six times during one oscillation period if the observer travels along with the density wave. Superharmonics oscillations of the local maxima of $\phi$ is indeed observed in experiments. As shown in Fig.~\ref{fig:wavenumber}, the distance between neighbouring density waves, $\lambda_c$, is larger than the diameter of their circular motion, i.e., $ d_w/\lambda_c\approx 0.5$, ensuring no intersection between their trajectories. Therefore, such superharmonics fluctuations suggest the out-of-alignment and realignment events between the flow and the hexagonal pattern.

The wavelength of the observed pattern, $\lambda_c$, is measured from the two-dimensional Fourier spectrum of the experimental images and averaged over one oscillation cycle, typified by Fig.~\ref{fig:wavenumber} inset. Data from the density-matched solution and the aqueous suspension are plotted in Fig.~\ref{fig:wavenumber}. 
Overall, $\lambda_c$ decreases with $\omega$, demonstrating a shift of dominance from long waves to short waves. 
Just above $\omega_c$, the relatively large error bars indicate that the structure is not well-ordered yet. For the intermediate $\omega>\omega_c$, $\lambda_c$ of the density-matched suspension is larger than the pure aqueous solvent, which indicates that the vertical migration of particles could modify the inhomogeneous pattern.
With increasing $\omega$, $\lambda_c$ decreases towards an asymptotic value where the aqueous suspension and the density-matched suspension match. The measured $\lambda_c$ is compared with linearly selected wave number $\lambda_m$, corresponding to the peak growth rate Fourier mode in the linearized model. The magnitude of $\eta_p$, bounded by $\rho\nu(\phi)=\rho\nu_{0}(\phi_J-\phi)^{-2}$, is critical for evaluating $\lambda_m$. From the perspective of linear analysis, $\eta_p\sim\rho\nu(\Phi)$ for the uniform state is implemented, and the resultant $\lambda_m$ is on the order of micrometers.
Nevertheless, as seen in Fig.~\ref{fig:uniform_densitywave}g, the local density, $\phi$, can be as high as 0.449 for the developed density waves, leading to viscosity proliferation locally.
The corresponding $\eta_p(\phi)$ would shift $\lambda_m$ towards longer wavelength. The evaluation of $\lambda_m$ with $\eta_p = 2800~\si{Pa \cdot s}$ is in reasonable agreement with $\lambda_c$ [Fig.~\ref{fig:wavenumber}]. It again indicates that the description of the fully-developed density pattern is beyond purely linear predictions~\citep{duru2002constitutive,Anderson1995}.

\section{Discussions and concluding remarks}
Our last remark is about the role of boundary confinement in the growth of density disturbances. 
Recent advances suggest that the steady shear-thickening state is a precursor to shear jamming~\citep{brown2012role,peters2016direct,mari2015nonmonotonic,singh2018constitutive}, where the dilation of the particle phase under shear is (partially) frustrated by the confining stress~\citep{Fall2008,brown2012role}.
The boundary confinement is thus considered essential. 
The open system studied so far is bounded by the air-suspension interfacial tension, $\Gamma$, and the curvature of the free surface, $\Delta h/\lambda_c^2$ (see Fig.~\ref{fig:uniform_densitywave}h). For the developed density waves, this confining pressure necessarily balances the onset stress of DST, $\Gamma \Delta h/\lambda_c^2\sim \tau^*$. Reducing $\Gamma$, e.g., by covering the suspension with a layer of 0.1\si{mm} thick silicone oil (\SI{10}{cSt}), decreases $\lambda_c$. 
On the other hand, confining the suspension with an acrylic plate, either comoving or fixed in the lab frame of reference, completely suppresses the density waves, though transient fluctuations of $\phi$ and thrust on the plate are present instead, as reported for shear-thickening suspensions in other configurations~\citep{lootens2003giant,rathee2020localized,rathee2017localized,hermes2016unsteady}. 
This dramatic contrast indicates that boundary confinement alters the underlying growth of disturbances. Indeed, long-lived inhomogeneities have only been reported near free surfaces previously~\citep{ovarlez2020density, hermes2016unsteady, Gauthier2023}.
We speculate that only density disturbances below a critical size may persist under rigid confinements. The existence of finite-size inhomogeneities may be readily manifested in experiments with the comoving top-plate confinement. 
In this configuration, a suspension stays uniform for $\omega>\omega_c$ as described above. However, soon after removing the confinement and restarting the oscillation at $\omega<\omega_c$, the inhomogeneous density profile appears and proliferates for a finite duration, then decay towards the uniform state (Movie 2). It implies that finite-size clusters already exist in the comoving-confinement configuration under shear~\citep{cheng2011imaging}. Though those clusters are too small to be accessible in our method, they may still be sufficiently large to trigger the transient growth even for $\omega<\omega_c$ after removing the rigid confinement.

The effect of boundary confinement discussed above is nontrivial for understanding the shear-thickening behavior. As evidenced by recent experiments with spatial resolution~\citep{rathee2017localized,rathee2020localized,saint2018uncovering,ovarlez2020density,gauthier2021new}, the shear-thickened state is intrinsically heterogeneous. 
In this paper, we have shown that the uniform state of a suspension under shear spontaneously breaks down due to the shear-thickening property. In addition, the growth and the manifestation of inhomogeneity highly depend on boundary conditions. 
With soft boundaries, such as the free surface experiments here, dilation is allowed to a certain extent. In this scenario, the inhomogeneity develops into a persistent density-wave state, where particles do not make long-lived contact. On the other hand, with rigid confinement, dilation is frustrated. When the dimension of a local high-density region is comparable to the gap between boundaries, a sudden rise in the stress response is expected~\citep{seto2013discontinuous,nabizadeh2022structure}. The intense stress compels particles into profound interactions, revealing features that might otherwise remain hidden, such as the role of particle adhesion~\citep{Gauthier2023} and the occurrence of hysteresis~\citep{galvez2017dramatic}. Furthermore, macroscopic clusters of particles exist only briefly under intense stress, leaving behind smaller ones that subsequently promote the reformation of high-density clusters. From a heterogeneous perspective, the overall stress response of a suspension in the conventional shear-thickened state (with rigid boundaries) is primarily governed by the increasingly prominent formation and collapse of these high-stress regions~\citep{rathee2017localized,rathee2020localized,van2023}.
The constitutive relation, calibrated with bulk rheology measurements, only captures these intermittent microscopic events on average. 
Therefore, the heterogeneous scenario fundamentally differs from the homogeneous one, similar to the discrimination between the parallel and serial relaxation schemes of glassy systems~\citep{Berthier2011}.
{Even though the mean-field theory could predict behaviors near the DST transition, such as the instability onset in this work, it may fail for the developed heterogeneous state, e.g., the effect of different boundary conditions. } Knowledge of microscopic/mesoscopic structures and dynamics are thus necessary to advance our understanding of the nature of shear-thickening suspensions.

\appendix

\section{Local particles fraction measurement}\label{appA}
\begin{figure}
	\centering
	\includegraphics[scale=0.8]{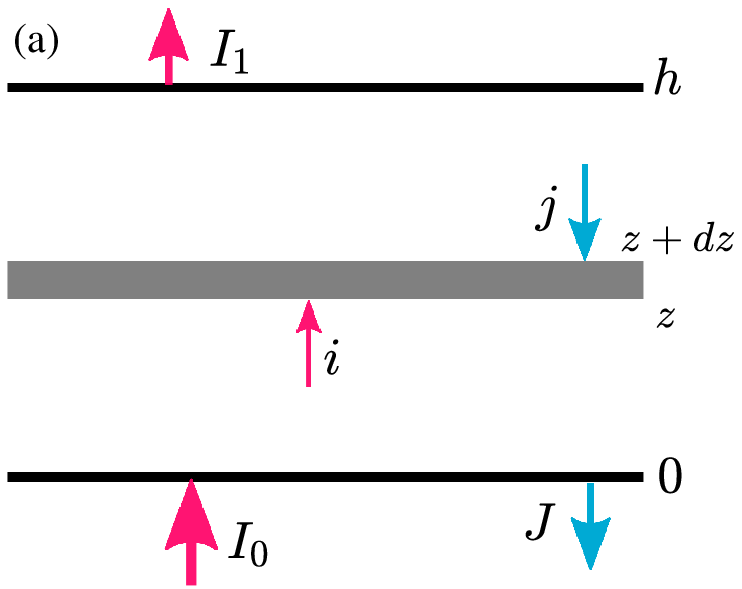}
	\includegraphics[scale=0.8]{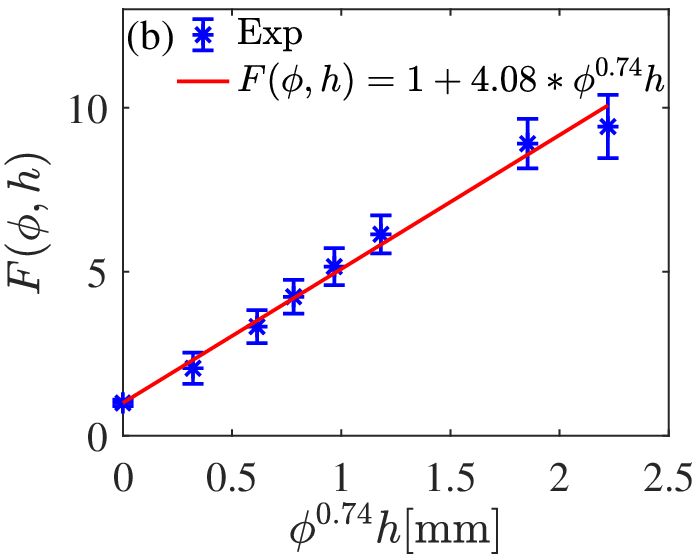}
	\caption{(a) At an arbitrary point inside the suspension layer, z. The intensity of the light backscattered by the layer at $z+\mathrm{d}z$ is denoted as $j$, and the intensity of the light going forward is denoted as $i$. $I_0$ is the total incident light intensity, and $I_1$ is the passing through light intensity, $h$ is the thickness of the suspension. (b) The relationship between light reduction ratio $F(\phi, h)$ and particle fraction $\phi$ and thickness $h$ of the cornstarch suspension.}
	\label{fig:multiple_scattering_theory}
\end{figure}

Only some is transmitted once light passes through a suspension, and the rest is either absorbed or scattered. To obtain a relation between the light transmission ratio and local density, we establish a one-dimensional model \citep{kubelka1931article}, described in Fig.~\ref{fig:multiple_scattering_theory}. An infinitesimal layer of the suspension absorbs and scatters a certain portion $Sdz+Rdz$ of the light of one unit of intensity passing through it, where $S$, the absorption coefficient and $R$, the scattering coefficient, are functions of local particle density $\phi$. The change of the forward and backward light intensity, $\mathrm{d}i$ and $\mathrm{d}j$, across the layer satisfy
\begin{eqnarray}
	di = -(S+R)idz+Sjdz
	\label{eq:mutiplescatter1}\\
	dj =  (S+R)jdz-Sidz.
	\label{eq:mutiplescatter2}
\end{eqnarray}
Note that the second terms on the right-hand side of Eqs.~\ref{eq:mutiplescatter1} and \ref{eq:mutiplescatter2} represent the contribution of back-scattered light. In other words, the back-scattered light intensity $j$ at $z$ is partially redirected forward by the layer at $z-\mathrm{d}z$ again through the back-scattering process and contributes to $i(z)$. This mechanism sets it apart from the Beer-Lambert law, where local light attenuation is completely lost in the final transmission.

For the suspension system studied here, cornstarch grains are white particles of irregular shapes, and the absorbing portion of the light is thus neglected ($S=0$). Therefore, the incident luminous flux equals the sum of transmission and reflection. Then the solution of Eqs.~\ref{eq:mutiplescatter1}-~\ref{eq:mutiplescatter2} with the corresponding boundary condition: $i(z=0) = I_0$ and $j(z=h)=0 $, is
\begin{equation}
	F(\phi,h) = \frac{I_0}{I_1} =1+R h.
	\label{eq:reduction}
\end{equation}
$R$ is regarded as a function of local density $\phi$, $R=a\phi ^{b}$, where $a$ represents the ratio of backscattering light per unit thickness, and $b$ is related to the particle shape. We fit the coefficient in Eq.~\ref{eq:reduction} using the light intensity attenuation data measured for uniform cornstarch suspensions, which gives $a=4.08~\si{mm^{-1}}$, $b=0.74$.

To obtain $\phi$ by Eq.~\ref{eq:reduction}, the local thickness $h$ is needed for the uneven surface at high excitation frequency (Fig.~\ref{fig:State_diagram}g), in addition to the intensity attenuation ratio of the backlight. The surface deformation is measured with a in-house-built high-speed laser profilometer. The surface deflection also introduces a focusing effect of the light, causing the trough to appear brighter. In our experiments, the maximum curvature of the deflection is approximately 0.04 \si{mm^{-1}} (Figure~\ref{fig:uniform_densitywave}g). The resultant intensity variant is less than 1\%, significantly lower than the observed value $\sim 20\%$. Therefore, we neglect this effect in the $\phi$ measurement.

Note that a uniform distribution of particles along the $z$ direction is assumed in the above analysis. In our experiments, particles may migrate to the surface of the suspension under certain conditions, which causes uneven distribution of particles and measurement error of $\phi$. 
To estimate the errors, we compare the average of $\phi$ measured via Eq.~\ref{eq:reduction} and the global particles fraction $\Phi$ in Table~\ref{tab:er}. The typical error is $3\%$ and slightly increases with $f$, confirming the conservation of mass.

We only considered the transmission flux in the model. In practice, the light is scattered in three-dimension. The transversal scattering causes blurring in the x-y plane. A 5mm thick suspension acts as a Gaussian blur kernel with a radius of 5.8 mm. The blurring radius increases approximately as the square of depth. Nonetheless, even with this blurring effect, the position of the maximum gradient of the intensity remains unchanged.

\begin{table}
	\begin{center}
		\def~{\hphantom{0}}
		\begin{tabular}{lccccccccc}
			
			\diagbox[dir=SE,innerwidth=1.5cm,height=1cm]{$\Phi$}{$\bar{\phi}/ \Phi$}{$f[\si{Hz}]$}&0.00&1.67&3.33&4.17&5.00&6.00&7.00&8.00&8.33\\ [3pt]
			0.355&1.00&0.99&1.00&1.01&1.02&1.03&1.02&1.02&1.01\\ 
			0.360&1.00&1.01&1.02&1.02&1.04&1.03&1.03&1.04&1.01\\ 
			0.370&1.00&1.00&1.00&0.99&1.00&1.01&0.99&0.98&0.98\\ 
			0.375&1.00&1.01&1.02&1.01&1.03&1.03&1.03&1.01&1.03\\ 
			0.380&1.00&1.00&1.01&1.01&1.00&1.01&1.00&1.05&1.02\\ 
			0.385&1.00&1.00&1.00&1.03&1.01&1.02&- &1.02&- \\ 
		\end{tabular}
		\caption{Evaluate the error of the density calculation method}
		\label{tab:er}
		\begin{tablenotes}
			\footnotesize
			\item[1] $\Phi$ is the given global density, $\bar{\phi}=\frac{1}{n}\Sigma\phi_i$ is the calculated average density, $f$ is the oscillation frequency.
		\end{tablenotes}
	\end{center}
\end{table}

\section{Linear stability analysis}\label{appB}

The stability analysis is performed with respect to the uniform state described by Eq.~\ref{eq.s_stokes} in the main text. A complex function describes the uniform flow, $U(z,t) = U_x + iU_y$. We use $h$, $\omega^{-1}$ and $A\omega$ as characteristic scales of length, time and velocity. The dimensionless solution of the flow field, $\hat{U} = U/(A\omega)$, is 
\[
\hat{U} = \frac{e^{-\hat{z}/l}}{1+e^{2(1+i)/l}}\left(e^{2(1+i)/l} + e^{2\hat{z}(1+i)/l}\right)e^{i(t-\hat{z}/l)},
\]
where $l=\sqrt{2\nu/\omega}/h$. The shear rate, $\dot{\gamma}=\lvert\partial U / \partial z\rvert$, and the shear stress, $\tau = \rho\nu\dot{\gamma}$, can be further calculated.
For simplicity, $\tilde{U} = A\omega\hat{U}(\hat{z}=1)$ and $\tilde{\tau} = \tau(z=0)$ are used to represent the mainstream. We give explicitly
\begin{equation}
	\tilde{\tau} = \rho\nu\dot{\gamma}(z=0) = \rho A h \omega^2 \frac{l}{\sqrt{2}} \sqrt{ \frac{\cosh(2/l) - \cos(2/l)}{\cosh(2/l) + \cos(2/l)}
	}
	\label{eq.tau}
\end{equation}

At given $\omega$ and $\Phi$, the flow of the uniform state and $\tilde{\tau}$ are thus fully described by $l\sim\sqrt{\nu}$. Since $\nu$ depends on shear stress in the shear-thickening constitutive relation, the values of $l$ and $\tilde{\tau}$ are mutually determined. We find their value by numerically converging the constitutive relation and the flow solution, i.e., Eq.~\ref{eq.tau} and 
\[\nu = \nu_0/(\Phi-\phi_J)^2,\quad \text{where } \phi_{J} = \phi_{0} - \exp(-\tau^*/\tilde{\tau})(\phi_{0}-\phi_{m}).\]
This concludes the calculation of the uniform state. Next, we simplify the two-phase model following the assumption stated in Section \ref{sec:onset} and then perform linear stability analysis on the reduced model with respect to the uniform state.

The one-dimensional version of Eq.1-2 in main text can be written as
\begin{equation}
	\frac{\partial \phi}{\partial t}+\frac{\partial({u}_p\phi)}{\partial\chi}=0
	\label{eq:1d_continue}
\end{equation}
and
\begin{equation}
	\phi \rho_{p} \left( \frac{\partial{u_p}}{\partial t} + u_p \frac{\partial u_p}{\partial \chi} \right) =C_d(\tilde{U}(\phi)-{u}_p)+\eta_p\frac{\partial ^{2} u_{p}}{\partial \chi^{2}}-\frac{\partial \Pi}{\partial \chi}
	\label{eq:1d_momentum}
\end{equation}
respectively. Note that $\chi$ represents the coordinate in the flow direction, different from $x$. As we focus our analysis along $\chi$, the time-dependent phase angle in $\tilde{U}$ is ignored. The Richardson-Zaki approximation is used for the drag coefficient,
\[	C_d = \frac{18\eta_f}{d^{2}}\frac{\phi}{(1-\phi)^5}.\]

The small-amplitude perturbations of $\phi$ and $u_p$ relative to $\Phi$ and $\tilde{U}$  are decomposed into a linear combination of Fourier modes, each has a complex growth rate $\sigma_k$.
\begin{gather*}
	\phi(\chi,t) = \Phi + \sum_{k}\hat{\phi}_{k} e^{ik \chi +\sigma_{k}t}\\
	u_p(\chi,t) = \tilde{U}(\Phi) + \sum_{k}\hat{u}_{k} e^{ik \chi+\sigma_{k}t}.
\end{gather*}
Equations~\ref{eq:1d_continue} and \ref{eq:1d_momentum} are then linearize in $\hat{\phi}_{k}$ and $\hat{u}_{k}$. The growth rate $\sigma_k$ satisfies the quadratic equations:
\begin{equation}
	(\sigma_k + i k \tilde{U})^2 + 
	\left(\frac{\eta_p k^2}{\rho_p \Phi} + \frac{C_d}{\rho_p \Phi} \right) (\sigma_k + i k \tilde{U} ) + 
	\frac{\Pi^{\prime}k^2}{\rho_p} + 
	i\frac{C_d \tilde{U}^{\prime}k}{\rho_p}=0.
	\label{eq:sigmak}
\end{equation}

The roots of Eq.~\ref{eq:sigmak} are 
\begin{gather}
	\operatorname{Re}(\sigma_{k})=\frac{-a\pm\sqrt{\frac{a^2+b+\sqrt{(a^2+b)^2+c^2}}{2}}}{2}
	\label{eq:resigmak}\\
	\operatorname{Im}(\sigma_{k})=\pm\frac{\sqrt{\frac{-a^2-b+\sqrt{(a^2+b)^2+c^2}}{2}}}{2} - k\tilde{U},
	\label{eq:imsigmak}
\end{gather}
where $a=\frac{\eta_p k^2}{\rho_p \Phi} + \frac{C_d}{\rho_p \Phi}>0$, $b=\frac{-4 \Pi^{\prime}k^2}{\rho_p}<0$, $c=\frac{4C_d \tilde{U}^{\prime}k}{\rho_p}$. The uniform flow ($\phi= \Phi$ and ${u_p}=\tilde{U}$) becomes unstable if $\operatorname{Re}(\sigma_{k})>0$. According to Eq.~\ref{eq:resigmak}, this criterion is equivalent to $c^2 + 4a^2b > 0 $: 
\begin{equation}
	1 + \frac{k^2 \eta_p}{C_d}<\frac{\phi \tilde{U}^{\prime}}{\sqrt{\Pi^{\prime}/\rho_p}},
	\label{eq:instabilitybound}
\end{equation}
where $\tilde{U}^{\prime}$ and $\Pi^{\prime}$ are the derivatives of $\tilde{U}$ and $\Pi$ with respect to $\phi$ at $\phi=\Phi$. Note that $\eta_p$ and $C_d$ are functions of $\phi$ in general. In Eq.~\ref{eq:instabilitybound}, $\eta_p(\Phi)$ and $C_d(\Phi)$ are understood, as their $\phi$ dependency leads to higher order corrections. In Equation~\ref{eq.instability} in the main text, the unity term on the left-hand side of Eq.~\ref{eq:instabilitybound} is replaced by a constant $\mathcal{C}$.

\begin{figure}
	\centering
	\includegraphics[width=0.49\textwidth]{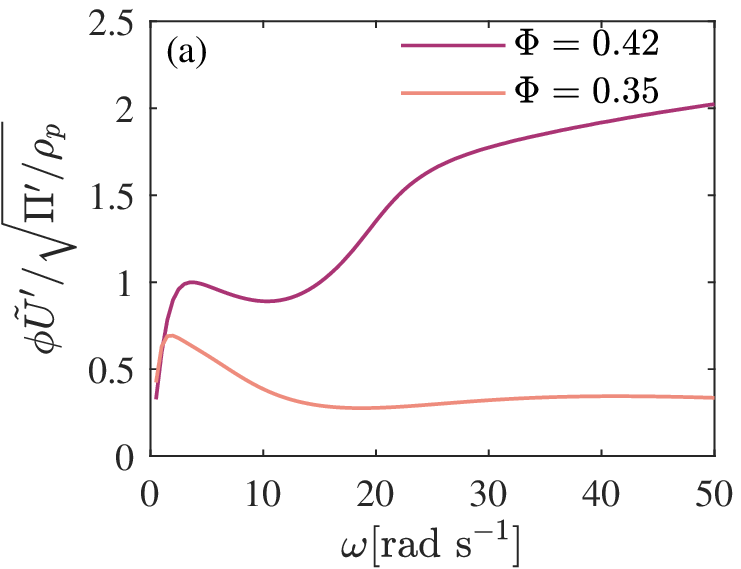}
	\includegraphics[width=0.49\textwidth]{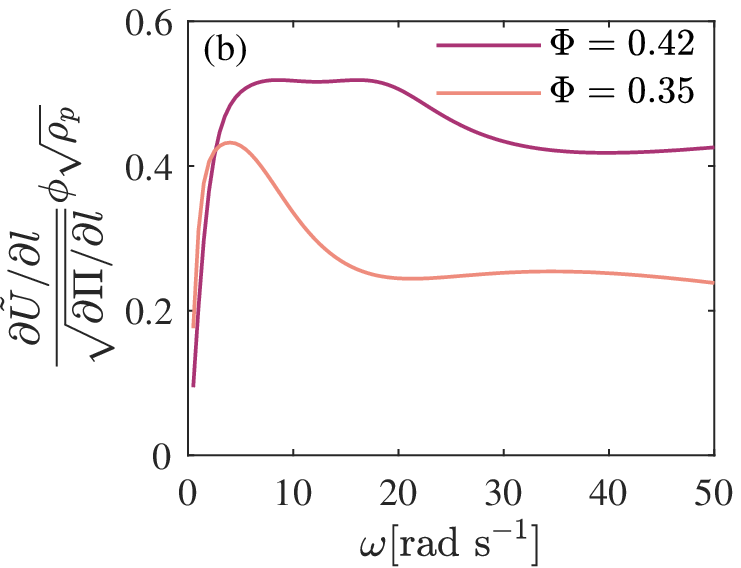}
	\includegraphics[width=0.49\textwidth]{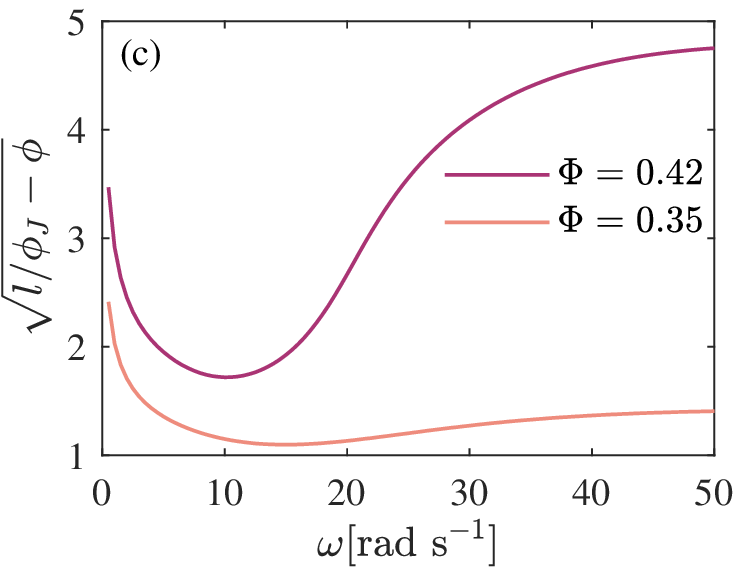}
	\includegraphics[width=0.49\textwidth]{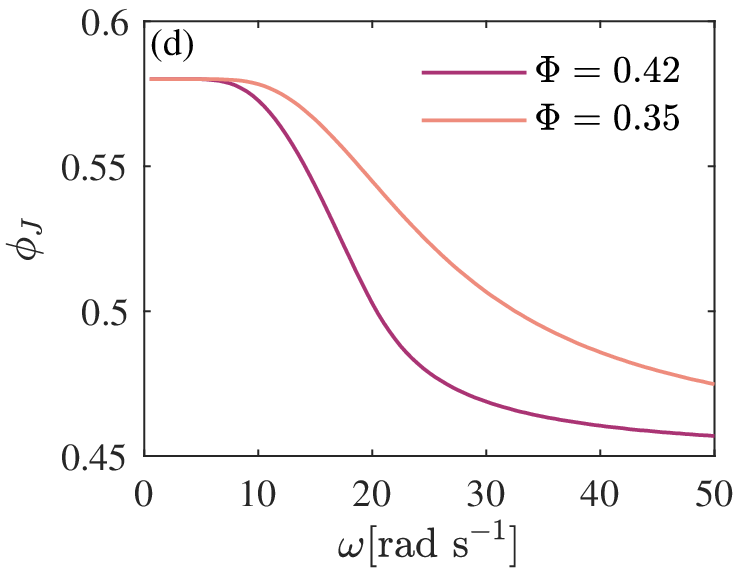}
	\caption{Theoretical computation of components of Eq.~\ref{eq:phimu}.}
	\label{fig:instabilitybound}
\end{figure}

The right-hand side of Eq.~\ref{eq:instabilitybound} is plotted versus $\omega$ in Fig.~\ref{fig:instabilitybound}(a). The parameters $\phi_0$, $\phi_m$, $\tau^{*}$, and $\nu_0$ used here are the same as in the main text. For $\Phi=0.42$, the ratio, $\frac{\phi \tilde{U}^{\prime}}{\sqrt{\Pi^{\prime}/\rho_p}}$, increases almost monotonically with $\omega$, and Equation 2 in the main text is satisfied beyond a critical value of $\omega_c$. However, for $\Phi=0.35$, this ratio increases slightly and then declines towards a constant smaller than 1.

\raggedbottom

To determine the key ingredient dominating the variation of ${\phi \tilde{U}^{\prime}}/{\sqrt{\Pi^{\prime}/\rho_p}}$, we we further decompose this ratio:
\begin{equation}
	\frac{\phi \tilde{U}^{\prime}}{\sqrt{\Pi^{\prime}/\rho_p}}=\left( \frac{\mathrm{d} {\tilde{U}}/{\mathrm{d} {l}}}{\sqrt{\mathrm{d} {\Pi}/\mathrm{d} {l}}}{\phi}\sqrt{\rho_p} \right )\sqrt{ l/(\phi_J-\phi)},
	\label{eq:phimu}
\end{equation}
where $\tilde{U}^{\prime}=\frac{\mathrm{d} \tilde{U}}{\mathrm{d} l} \frac{l}{\phi_J-\phi}$, ${\Pi}^{\prime}=\frac{\mathrm{d}{\Pi}}{\mathrm{d} l} \frac{l}{\phi_J-\phi}$ are understood.
The two terms on the right hand side of Eq.~\ref{eq:phimu} are referred to as $o_1 = \frac{\mathrm{d}{\tilde{U}}/{\mathrm{d}{l}}}{\sqrt{\mathrm{d}{\Pi}/\mathrm{d}{l}}}{\phi}\sqrt{\rho_p}$ and $o_2 = \sqrt{l/(\phi_J-\phi)}$. As illustrated in Fig.~\ref{fig:instabilitybound}(b), the term $o_1$ initially increases with $\omega$ but soon saturates. In contrast, the dramatic increasing of $o_2$ (in particular, of $l$) dominates the ratio, $\frac{\phi \tilde{U}^{\prime}}{\sqrt{\Pi^{\prime}/\rho_p}}$, beyond $\omega\approx 10~\si{rad~s^{-1}}$, as shown in Fig.~\ref{fig:instabilitybound}(c-d).

\section{Rheology measurement on the cornstarch suspension}
\label{sec:rheo}
In the constitutive relation, $\nu_0,\tau^{*},\phi_0,\phi_m$ are parameters. We determine these parameters via the rheology properties of the aqueous suspension of cornstarch used in the experiment ( Anton Paar 302 ). The rheological data fitting procedure is as follows: The low viscosity branch ( frictionless branch ) is first fitted with $\eta_s=\nu_0 \rho_s (\phi_0 - \phi)^{-2}$ ($\rho_s$ is the solvent density), with $\nu_0$ and $\phi_0$ as the fitting parameters, and yields $\nu_0\in [0.57~ 0.95] 
\times 10^{-6}~\si{m^{2}/s}$ and $\phi_0\in[0.563~0.585]$. The high viscosity branch  ( frictional branch ) is then fitted with  $\eta_s=\nu_0 \rho_s (\phi_m - \phi)^{-2}$, using the previous estimation of $\nu_0$ and leaving $\phi_m$ as the only fitting parameter, which yields $\phi_m \in [0.443~0.458]$. Once the values of $\nu_0$, $\phi_0$ and $\phi_m$ are set, we determine the value of $\tau^{*}$ by fitting the full rheograms $\eta_s(\tau)$ with Wyart-Cates model: $\eta_s(\phi)=\nu_0 \rho_s (\phi_J (\tau) -\phi)^{-2}$, with $\phi_J(\tau) =\phi_0 -\exp^{-\tau^{*}/\tau}(\phi_0 -\phi_m)$. The best fit, shown in Fig.S6, gives for $\tau^{*} \in [2.6~8]\si{Pa}$, where the fitted value decreases with $\Phi$. $\tau^{*}$ represents the critical stress required to overcome the inter-particle repulsive force and activate frictional contacts between particles, and the average of its fitted value for relatively high $\Phi$ is taken. Based on the analysis above, $\nu_0=0.95 \times 10^{-6}~\si{m^{2}/s}$, $\phi_m=0.45$, $\tau^{*}=\SI{3.7}{Pa}$ and $\phi_0=0.58$ are used in the main text.
\begin{figure}
	\centering
	\includegraphics[scale=0.6]{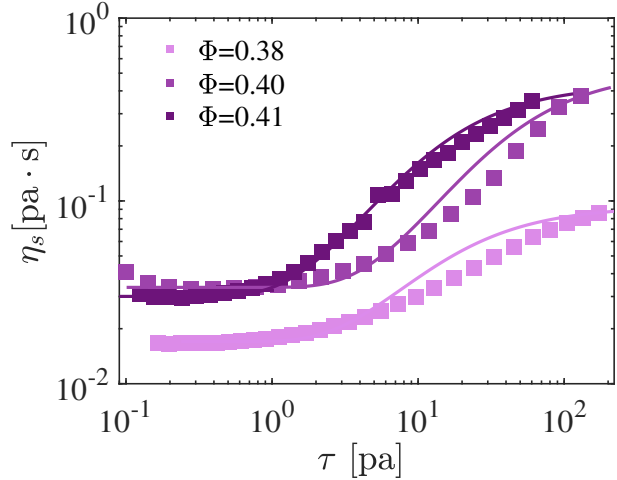}
	\label{fig:rheology}
	\caption{Rheological data for $\Phi=0.38,\Phi=0.40,\Phi=0.41$ (points), and the fitting (solid lines) with Wyart-Cates relation. The data is presented in the suspension viscosity $\eta_s$ and the shear stress $\tau$.}
\end{figure}
\bibliography{wave_JFM}
\end{document}